\documentclass[preprintnumbers]{PoS}
\usepackage{amsmath,amssymb}
\usepackage[usenames,dvipsnames]{xcolor}
\usepackage{graphicx}
\usepackage{epsfig}
\usepackage{rotating}
\usepackage[normalem]{ulem}

\newcommand\bef{\begin{figure}}
\newcommand\eef[1]{\label{fg:#1}\end{figure}}
\newcommand\besf{\begin{subfigure}}
\newcommand\eesf[1]{\label{sfg:#1}\end{subfigure}}
\newcommand\beq{\begin{equation}}
\newcommand\eeq[1]{\label{eq:#1}\end{equation}}
\newcommand\beqa{\begin{eqnarray}}
\newcommand\eeqa[1]{\label{eq:#1}\end{eqnarray}}
\newcommand\bet{\begin{table}}
\newcommand\eet[1]{\label{tb:#1}\end{table}}
\newcommand\best{\begin{subtable}}
\newcommand\eest[1]{\label{stb:#1}\end{subtable}}
\newcommand\betb{\begin{center}\begin{tabular}}
\newcommand\eetb{\end{tabular}\end{center}}
\newcommand\beit{\begin{itemize}}
\newcommand\eeit{\end{itemize}}

\newcommand\fgn[1]{Figure \ref{fg:#1}}

\newcommand\scn[1]{Section \ref{sec:#1}}

\newcommand\tbn[1]{Table \ref{tb:#1}}

\newcommand{\nn}{\nonumber}

\newcommand\dadt[4]{[\bar{#1}\bar{#2}]_{3_c}[#3 #4]_{\bar{3}_c}}
\newcommand\dads[4]{[\bar{#1}\bar{#2}]_{\bar{6}_c}[#3 #4]_{6_c}}
\newcommand\dad[4]{[\bar{#1}\bar{#2}][#3 #4]}

\title{X(3872) and Y(4140) using diquark-antidiquark operators with lattice QCD}

\ShortTitle{X(3872) and Y(4140) using diquark-antidiquark operators with lattice QCD}

\author{\speaker{M.\ Padmanath}\\
        Institute of Physics, University of Graz, A-8010 Graz, Austria.\\
        E-mail: \email{padmanath.madanagopalan@uni-graz.at}}

\author{C.\ B.\ Lang\\
        Institute of Physics, University of Graz, A-8010 Graz, Austria.\\
        E-mail: \email{christian.lang@uni-graz.at}}

\author{S.\ Prelovsek\\
        Department of Physics, University of Ljubljana, Jadranska 19, 1000 Ljubljana, Slovenia \& \\
        Jozef Stefan Institute, Jadranska 19, 1000 Ljubljana, Slovenia \\
        Theory Center, Jefferson Lab, 12000 Jefferson Avenue, Newport News, VA 23606, USA \\
        E-mail: \email{sasa.prelovsek@ijs.si}}

\abstract{We discuss a recent lattice study of charmonium-like mesons with 
$J^{PC}=1^{++}$ and three quark contents $\bar cc\bar du$, $\bar cc(\bar uu + \bar dd)$ 
and $\bar cc\bar ss$, where the latter two can mix with $\bar cc$. In this 
quantum channel, the long known exotic candidate, X(3872), resides. This simulation 
employs $N_f=2$, $m_\pi=266~$MeV and a large basis of $\bar cc$, two-meson and 
diquark-antidiquark interpolating fields, with diquarks in both anti-triplet and 
sextet color representations. It aims at the possible signatures of four-quark 
exotic states. Along the way, we discuss the relations between the 
diquark-antidiquark operators and the two-meson operators via the Fierz transformations.}

\FullConference{The 33rd International Symposium on Lattice Field Theory\\
		14 -18 July 2015\\
		Kobe International Conference Center, Kobe, Japan*}

\begin{document}

\section{Introduction\label{sec:Intro}}

Existence of hadrons with exotic flavor quantum numbers and exotic nature has been an interesting 
open question in hadron spectroscopy. Studies in this direction achieved 
a boost with the discovery of charged resonances $Z_c(3900)^+$ \cite{Ablikim:2013mio} and 
$Z(4430)^{\pm}$ \cite{Choi:2007wga,Mizuk:2009da} that confirms the existence of hadrons 
composed of two quarks and two antiquarks. Many of the other exotic excitations in the charmonium 
sector, classified as so-called the XYZ mesons, also appear to have significant four-quark Fock components.

In this talk, we present the results from our recent lattice investigation of charmonium 
spectrum in the $J^{PC}=1^{++}$ quantum channel with three quark contents: $\bar cc \bar du$, 
$\bar cc(\bar uu+\bar dd)$ and $\bar cc \bar ss$, where the latter two channels have $I\!=\!0$ 
and can mix with $\bar cc$ \cite{Padmanath:2015era}. These calculations were aimed at a first principles study of 
$X(3872)$ and $Y(4140)$ and other possible hadrons in these channels.

Experimentally, the quantum numbers of $X(3872)$ have been confirmed to be $J^{PC}=1^{++}$,
while its isospin, $I$, remains unsettled. This is mainly due to its nearly equal branching fraction 
to either isospin decay channel, $I=0$ ($X(3872) \rightarrow J/\psi\, \omega$) and $I=1$ 
($X(3872) \rightarrow J/\psi \,\rho$) \cite{Agashe:2014kda}, and lack of evidence for existence 
of any charged partner states \cite{Aubert:2004zr}. Other XYZ candidates with $C\!=\!+1$ that 
could possibly have $J^{PC}=1^{++}$ include $X(3940)$ \cite{Abe:2004zs}, $Z(4050)^{\pm}$ 
\cite{Mizuk:2009da} and $Z(4250)^{\pm}$ \cite{Mizuk:2009da}. The signatures for the $Y(4140)$ 
with charge parity $C\!=\!+1$ in the $J/\psi \phi$ invariant mass \cite{Aaltonen:2009tz},
indicates the existence of exotic hadrons with hidden strangeness. However, the quantum numbers 
for most of these excitations are undetermined. A detailed review on these can be found in 
Refs. \cite{Olsen:2014qna}.

Several theoretical calculations based on phenomenology have been performed, which 
suggests a variety of interpretations for these exotic observations : mesonic 
molecules, hybrid mesons, tetraquarks, cusp phenomena, etc. Detailed reports on these 
investigations can be found in the Refs. \cite{Olsen:2014qna}. Lattice QCD promises a first 
principles approach to study these systems so as to establish their fundamental nature.
A lattice candidate for $X(3872)$ with $I\!=\!0$ was first reported in Ref. \cite{Prelovsek:2013cra},
where a combination of $\bar cc$ as well as  $D\bar D^*$ and $J/\psi \omega$ interpolators was used.
Recently, another lattice investigation using the HISQ action also reported a lattice candidate 
for $X(3872)$ using $\bar{c}c$ and $D\bar{D}^*$ interpolating fields \cite{Lee:2014uta} 
supporting the previous observation. There has been no evidence from lattice studies supporting 
existence of the $Y(4140)$ resonance, even though there have been lattice calculations of 
$J/\psi \phi$ scattering in search of the $Y(4140)$ resonance \cite{Ozaki:2012ce}.

In this lattice investigation, we make a dynamical study involving diquark-antidiquark interpolators
along with several two-meson and $\bar cc$ kind of interpolators, so as to explore the significance 
of tetraquark Fock components in the charmonium spectrum with $J^{PC}=1^{++}$. We consider the color 
structures ${\mathcal G}=\bar 3_c,6_c$  for the diquarks. This study addresses the following questions: 
What are the effects of diquark-antidiquark interpolators on the established lattice candidate for 
$X(3872)$ and the charmonium spectrum itself? Do we observe additional levels for charged or neutral 
$X(3872)$ with $I\!=\!1$? Do we see additional levels for $Y(4140)$ using operators with hidden 
strangeness? Do we find signatures for other possible exotic states in the channels being probed?

The paper is organized as follows. \scn{method} discusses the details of the lattice ensemble, the interpolators
and the technologies used in this investigation. In \scn{Fierz}, the relation between 
diquark-antidiquark and two-meson interpolators via Fierz transformations is discussed. In \scn{Results} and \scn{Conc}
the results and conclusions are presented. 


\section{Lattice methodology\label{sec:method}}

$N_f\!=\!2$ dynamical gauge configurations with $m_{\pi}\!\simeq \!266$ MeV \cite{Hasenfratz:2008ce} 
were used in these calculations. Other relevant details of the gauge ensemble are described in 
\tbn{latpar}. The sea and valence quarks were realized with the tree-level improved Wilson clover action. The absence 
of a strange sea prevents $\bar cc\bar ss$ intermediate states  in the  $\bar cc(\bar uu+\bar dd)$ sector, 
in accordance with treating these  two $I=0$ sectors separately in our study. 
Charm quarks are tuned by equating the spin averaged kinetic mass of the $1S$ charmonium to its physical 
value. We quote our spectrum in terms of $E_n=E^{lat}_n-m_{s.a.}^{lat}+m_{s.a.}^{exp}$ ($m_{s.a.} = \frac14(m_{\eta_c}+3m_{J/\psi})$ ), 
which will be compared with the experiments. 

\bet[tbh]
\centering
\betb{ccccccccccccc}
\hline
Lattice size     & $\kappa$ & $\beta$ &  $N_{\mathrm{cfgs}}$ &  $m_\pi$ [MeV] &    $a$ [fm]    & $L$ [fm]  \\\hline
$16^3 \times 32$ & $0.1283$ &   7.1   &        280           &    266(3)(3) & $0.1239(13)$ &  1.98     
\\\hline
\eetb
\caption{Details of the gauge field ensemble used.}
\eet{latpar} 

Altogether 22 interpolators, including $\bar cc$, two-meson and diquark-antidiquark ($\dad{c}{q}{c}{q}$) interpolators, 
with $J^{PC}=1^{++}$ ($T_1^{++}$ irrep of octahedral group, $O_h$) and total momentum zero were constructed for 
the three cases of our interest. Using these interpolators, the relevant non-interacting two-meson levels with 
$J^{PC}=1^{++}$ and total momentum zero
are 
\beit
\item ~~$I = 0$; ~$\bar cc(\bar uu+\bar dd)$ and $\bar cc$; ~$E\lesssim4.2~$GeV \\
$ ~~D(0)\bar D^*(0), ~J/\psi(0)\omega(0), ~D(1)\bar D^*(-1), ~J/\psi(1)\omega(-1), ~\eta_c(1)\sigma(-1), ~\chi_{c1}(0)\sigma(0)\;$
\item ~~$I = 1$; ~$\bar cc \bar du$ ; ~$E\lesssim4.2~$GeV \\
$ ~~D(0)\bar D^*(0), ~J/\psi(0)\rho(0), ~D(1)\bar D^*(-1), ~J/\psi(1)\rho(-1), ~\chi_{c1}(1)\pi(-1), ~\chi_{c0}(1)\pi(-1)\; $
\item ~~$I = 0$; ~$\bar cc \bar ss$  and $\bar cc$; ~$E\lesssim4.3~$GeV \\
$ ~~D_s(0)\bar D_s^*(0), ~J/\psi(0)\phi(0), ~D_s(1)\bar D_s^*(-1), ~J/\psi(1)\phi(-1)\;.$
\eeit

The flavor sectors $\bar cc(\bar uu+\bar dd)$ and $\bar cc \bar ss$ are considered separately in this calculation. 
We assume that either of these sectors have negligible coupling to two-meson levels from other. It is to be noted that that $Y(4140)$ 
has been experimentally observed only in the $J/\psi \phi$ final state with valence strange content, but it has not been 
observed in $D\bar D^*$ and $J/\psi \omega$ final states. With this assumption the resulting spectrum would be 
less dense, making the identification of the eigenstates easier. Furthermore, we assume that the valence strange content 
could uncover hints on the existence of the charm-strange exotics, if they exists, although the dynamical strange quarks 
are absent in this ensemble.

We show the two-meson non-interacting energies by the horizontal lines in our plots. They are calculated from 
energies of single hadrons determined on the same ensemble \cite{Lang:2011mn,Lang:2014tia,Mohler:2012na} 
using $E^{n.i.}_{M_1(\mathbf{n})M_2(-\mathbf{n})}= E_1(p) + E_2(p),\ p=\frac{2\pi |\mathbf{n}|}{L},\  \mathbf{n}\in N^3$. 
Some hadrons in the list above (e.g. $\rho$ and $\sigma$) are resonances and a proper simulation should consider
three meson interpolators. Such a calculation has not been performed in practice yet. 
Hence we follow an approximation in which the energy of the low-lying state from single hadron correlations are used 
to determine the two-meson non-interacting levels. We also exclude the relevant non-resonant three meson levels in 
the energy range of our 
interest, as we expect these three meson levels would be absent without explicit inclusion of respective interpolators
in the operator basis.

We compute the full coupled correlation functions with our interpolator set using the `Distillation' method \cite{Peardon:2009gh}.
In this study, we exclude Wick contraction diagrams, where the charm quark lines are disconnected between the source and sink, 
and we take into account all other Wick contractions.
The eigensystem are extracted using the well-established generalized eigenvalue problem \cite{Michael:1985ne} and the energies
are extracted asymptotically from two-exponential fits to the eigenvalues. All quoted statistical errors are obtained using single 
elimination jackknife analysis. The existence of possible exotic states is studied by analyzing the number of energy 
levels, their positions and overlaps with the considered lattice operators $\langle \Omega |O_j|n\rangle$. Based on experience
of identifying additional levels corresponding to $\rho$ \cite{Lang:2011mn}, $K^*(892)$ \cite{Prelovsek:2013ela}, $D^*_0(2400)$ 
\cite{Mohler:2012na}, $K_0^*(1430)$ \cite{Dudek:2014qha} and $X(3872)$ \cite{Prelovsek:2013cra}, we expect an additional 
energy level if an exotic state is of similar origin.


\vspace{-0.2cm}
\section{Fierz relations \label{sec:Fierz} }
The diquark-antidiquark operators  $\dadt{c}{q}{c}{q}$ and $\dads{c}{q}{c}{q}$ are linearly related to the 
two-meson currents $(\bar cc)_{1_c}(\bar qq)_{1_c}$ and $(\bar cq)_{1_c}(\bar qc)_{1_c}$ via Fierz rearrangement
\cite{Nieves:2003in}. We show an example of such a relation between one of our $\dad{c}{q}{c}{q}$ operators
and two-meson operators in eq. \ref{opfq1}, which is derived for local quarks. The Fierz relations suggest that $O^{4q}$ and $O^{MM}$ are 
linearly dependent, even though our quark fields are smeared and mesons are projected to definite momentum. 
Note that the first and second terms in the Fierz expansion (\ref{opfq1}) represent $D\bar D^*$, while the 
seventh term is similar to the $\chi_{c1} ~ \sigma$ for $I=0$. Hence we expect significant correlations between 
these operators. 
\beqa
O_{3_c}^{4q} &=& [\bar{c}~C\gamma_5~\bar{u}]_{\bar{3}_c}[c~\gamma_{i}C~u]_{3_c} + [\bar{c}~C\gamma_{i}~\bar{u}]_{\bar{3}_c}[c~\gamma_5C~u]_{3_c} + \{u\rightarrow d\} \label{opfq1} \\
&=& -\frac{(-1)^{i}}2 \{ ~(\bar{c}~\gamma_5~u)(\bar u~\gamma_{i}~c) -~(\bar{c}~\gamma_{i}u)(\bar u~\gamma_5~c) +~(\bar{c}~\gamma^{\nu}\gamma_5~u)(\bar u~\gamma_{i}\gamma_{\nu}~c) |_{i\ne\nu} \nn \\ 
&& ~-~(\bar{c}~\gamma_{i}\gamma_{\nu}~u)(\bar u~\gamma^{\nu}\gamma_5~c) |_{i\ne\nu} \} +\frac{(-1)^{i}}2\{~(\bar{c}~c)(\bar u~\gamma_{i}\gamma_5~u) +~(\bar{c}~\gamma_{i}\gamma_5~c)(\bar u~u)  \nn \\
&& -~(\bar{c}~\gamma^{\nu}c)(\bar u~\gamma_{i}\gamma_{\nu}\gamma_5~u) |_{i\ne\nu} ~-~(\bar{c}~\sigma^{\alpha\beta}~c)(\bar u~\sigma_{\alpha\beta}\gamma_{i}\gamma_5~u)|_{i\ne(\alpha<\beta)}\} +~ \{u\rightarrow d\} \nn 
\eeqa{opfq12}

\section{Results\label{sec:Results}}

\begin{figure}[tbh]
\parbox{.45\linewidth}{
\centering
\includegraphics[scale=0.72]{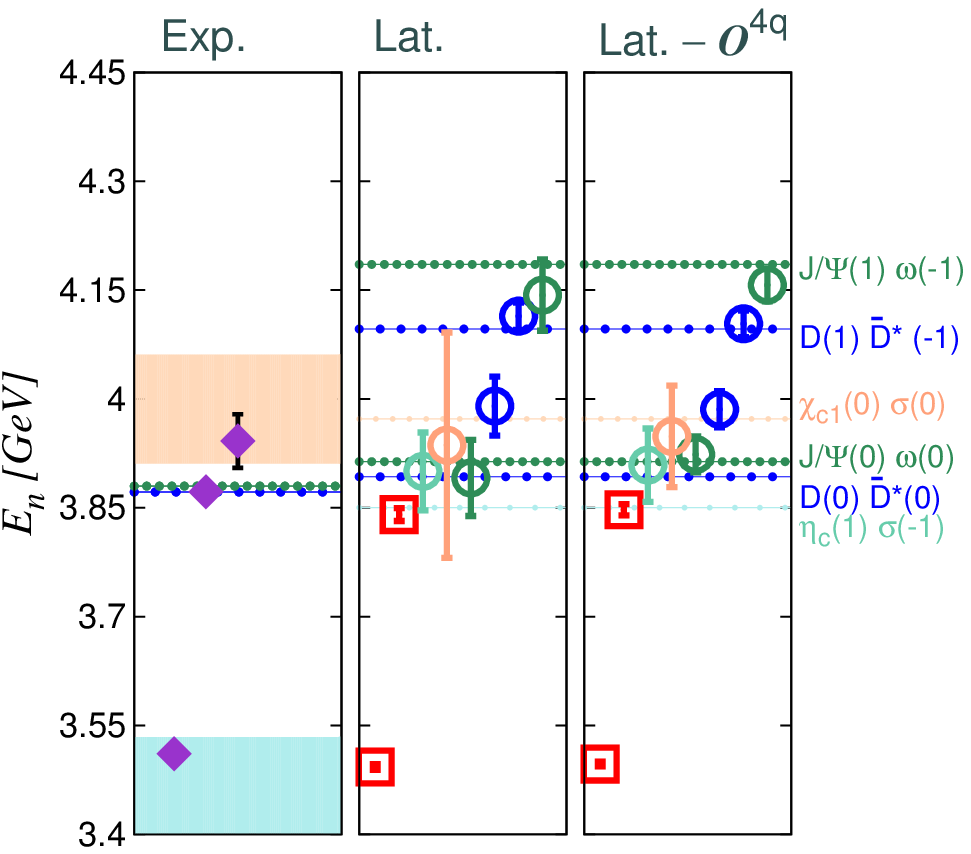}\\
(a) $I = 0: ~\bar cc(\bar uu+\bar dd)\ \&\ \bar cc$}
\hspace{0.9cm}
\parbox{.45\linewidth}{ 
\centering
\includegraphics[scale=0.72]{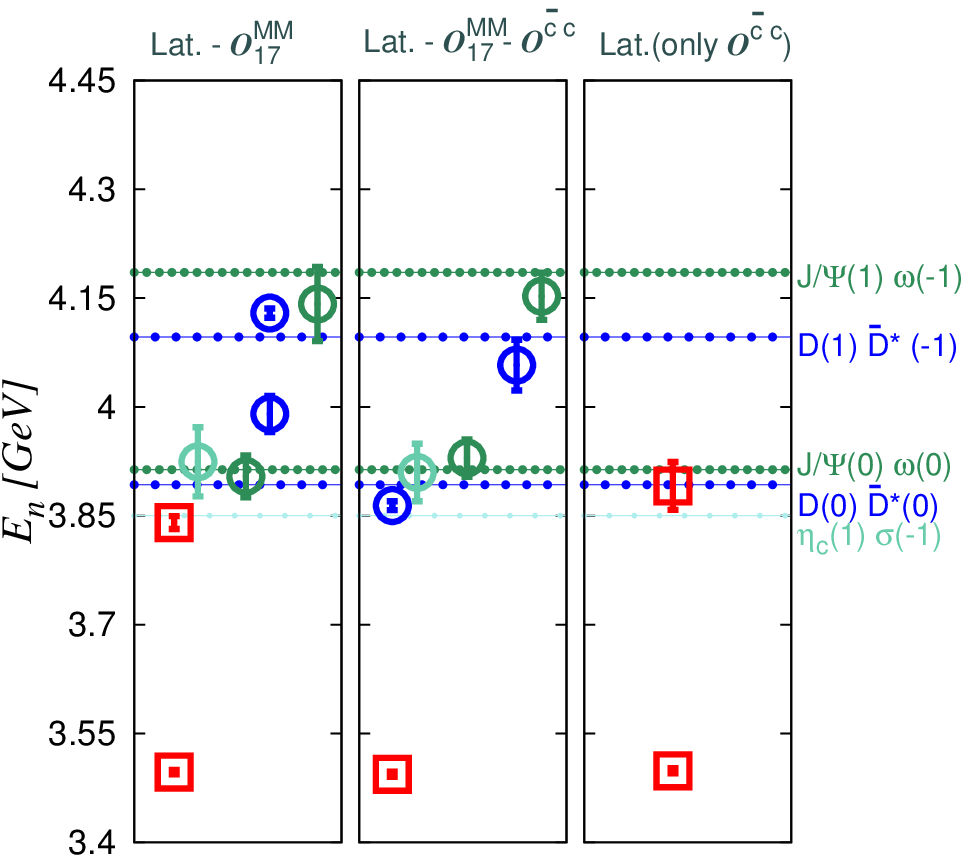}\\
(b) $I = 0: ~\bar cc(\bar uu+\bar dd)\ \&\ \bar cc$}
\caption{The I=0 spectrum with $J^{PC}=1^{++}$ with $u/d$ valence quarks. The energies are shown as 
$E_n=E^{lat}_n-m_{s.a.}^{lat}+m_{s.a.}^{exp}$. The two-meson non-interacting levels and experimental thresholds 
are displayed as horizontal lines, where the colored bands indicate the $\sigma$ width.
(a) The middle block shows the discrete spectrum determined from our lattice simulation, while the right-hand block
shows the spectrum we obtained with the $\dad{c}{q}{c}{q}$ operators excluded from our analysis. 
The left-hand block shows the physical thresholds and possible experimental candidates $\chi_{c1}$, 
$X(3872)$ and $X(3940)$. The violet error-bars for experimental candidates 
show the uncertainties in the energy and the black error-bars show its width. 
(b) The left block shows the spectrum from interpolator basis containing all kinds of operators.
The middle block shows the spectrum after excluding $\bar cc$ kind of operators. The right hand side block 
is the spectrum extracted purely from $\bar cc$ kind of operators. The $O_{17}^{MM}=\chi_{c1}(0)\sigma(0)$ is 
excluded from the basis to achieve better signals and clear comparison.}
\label{fg:primresl}
\end{figure}

\fgn{primresl} and \fgn{primres} show the discrete energy spectrum, which is the main result from this lattice 
calculation. The lattice energy levels are indicated by circles, while the horizontal lines indicate the energies of two meson states in the non-interacting limit. We identify the eigenstates that 
have dominant overlaps with two-meson scattering interpolators based on the spectral overlaps $\langle0|O_i|n\rangle$ and additional criteria described in Ref. \cite{Padmanath:2015era}. The corresponding circles are represented with the same colors as the lines for the non-interacting two meson states. The remaining states, that are not attributed to the two-meson states, are represented by red squares. 
These figures also compare the spectra obtained from two interpolator sets, one with the diquark-antidiquark operators 
and other without. In all three flavor sectors, we see almost 
negligible effect of $\dad{c}{q}{c}{q}$ on the low lying states, while we do observe an improvement in the signals for higher 
lying states in the basis without $\dad{c}{q}{c}{q}$.

\fgn{primresl} shows the I=0 charmonium spectrum with $J^{PC}=1^{++}$ and $u/d$ valence quarks. We identify the levels related to $X(3872)$ as $n=2$ (red squares) and $n=6$ (blue circle). One of 
the two levels remains absent when $D\bar D^*$ and $O^{4q}$ are used and $O^{\bar cc}$ is not, as is evident from the 
first and second panel from left of \fgn{primresl}(b). This indicates that the importance of $\bar cc$ interpolators 
for lattice candidate of $X(3872)$, while the $\dad{c}{q}{c}{q}$ structure alone does not produce it. Furthermore, 
it also indicates the significance of $\bar{c}c$ and $D\bar D^*$ operators in determining the position of these two 
levels, while the $O^{4q}$ doesn't have any significant implications on them. We extract the $DD^*$ scattering matrix S(E) at 
two energy values $E_{n=2,6}$ using L\"uscher's relation. The scattering matrix is interpolated near the threshold and a pole just 
below threshold is found \cite{Padmanath:2015era}. The results indicate a shallow 
bound state immediately below $D\bar D^*$ threshold, interpreted as experimentally observed $X(3872)$. The extracted 
mass and binding energy of $X(3872)$ indicate that it is insensitive to the $\dad{c}{q}{c}{q}$ interpolators. The mass of 
$X(3872)$ was determined along these lines for the first time in Ref. \cite{Prelovsek:2013cra}, where this channel 
was studied in a smaller energy range on the same ensemble without $\dad{c}{q}{c}{q}$ interpolators. All other extracted 
levels are identified with different two meson scattering channels.

\begin{figure*}[tbh]
\parbox{.45\linewidth}{
\centering
\includegraphics[scale=0.72]{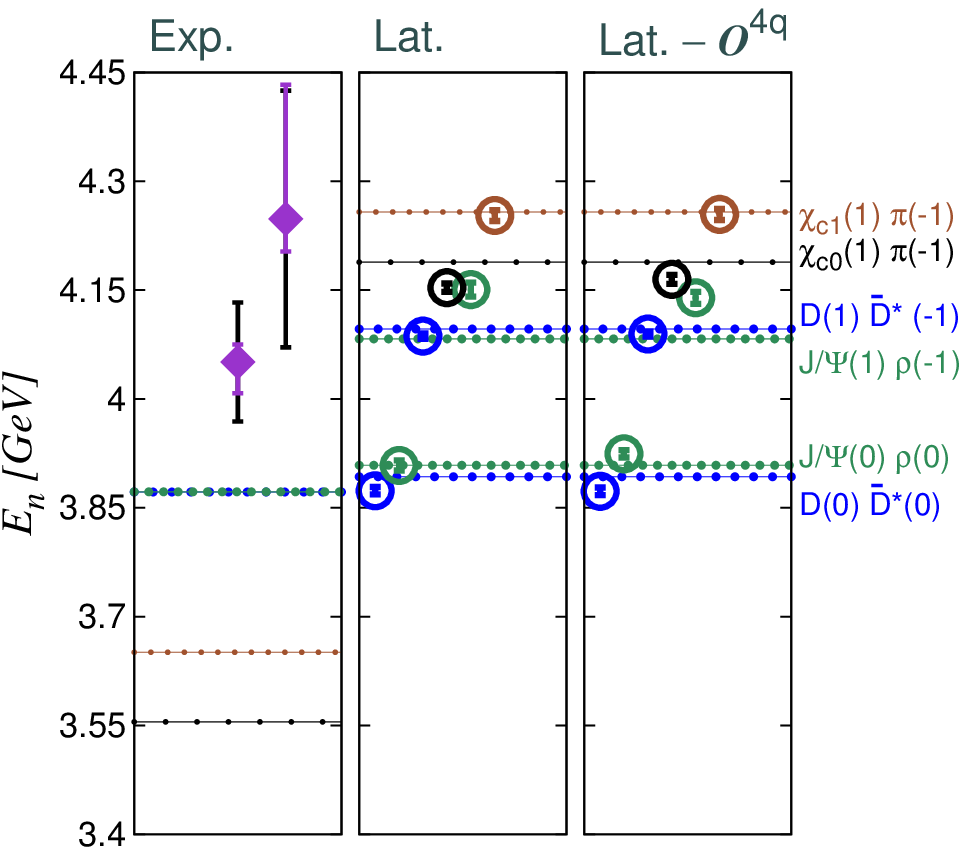}\\
(b) $I = 1: ~\bar cc\bar ud$}
\hspace{0.9cm}
\parbox{.45\linewidth}{ 
\centering
\includegraphics[scale=0.72]{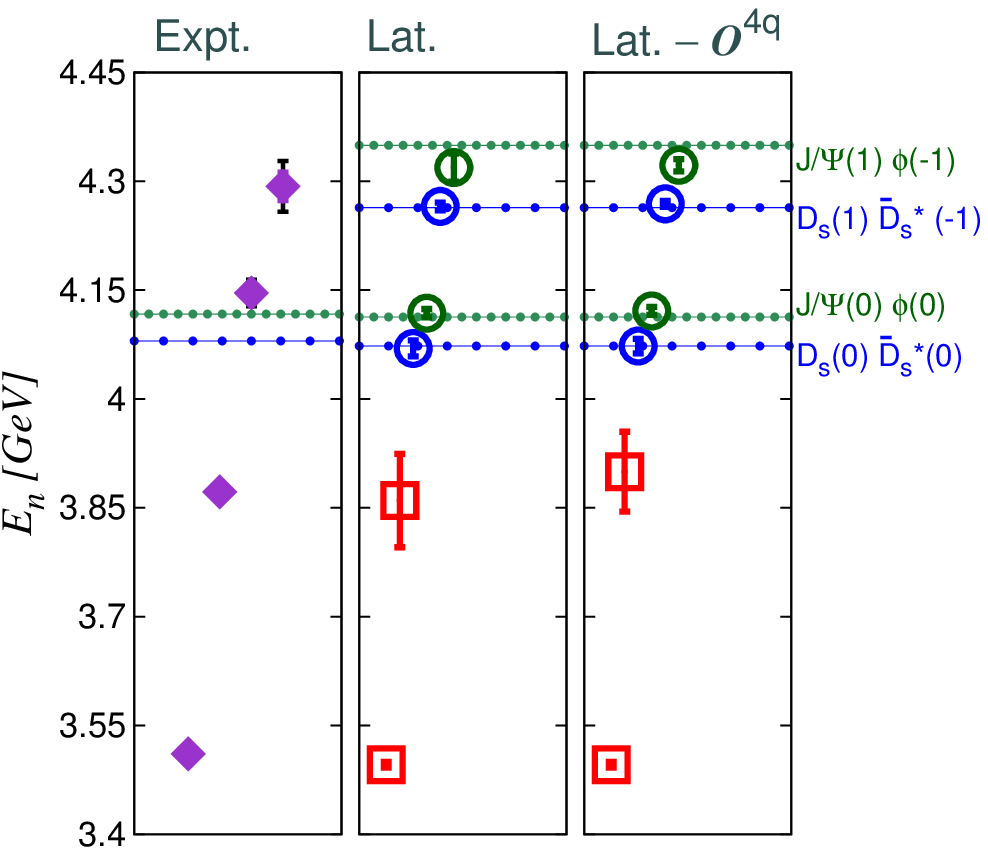}\\
(c) $I = 0: ~\bar cc\bar ss\ \&\ \bar cc$}
\caption{(a) The I=1 spectrum with $J^{PC}=1^{++}$ and (b) the I=0 spectrum with $J^{PC}=1^{++}$ with hidden 
strange valence quarks. The experimental candidates shown are (a) $Z_c^+$(4050) and $Z_c^+$(4250) and (b) $\chi_{c1}$, 
$X(3872)$, $Y(4140)$ and $Y(4274)$. For further details see Figure 1.}
\label{fg:primres}
\end{figure*}

\fgn{primres}(a) shows the I=1 spectrum with $J^{PC}=1^{++}$ and quark content $\bar cc \bar du$. All the eigenstates have 
dominant overlap with the two-meson interpolators. The spectrum shows very little influence on the inclusion 
of $\dad{c}{q}{c}{q}$, which is evident from \fgn{primres}(a). Our results do not give evidence for a charged 
or neutral $X(3872)$ with $I=1$ or other charged exotic mesons like $Z_c(4050)^+$ and $Z_c(4250)^+$.

\fgn{primres}(b) shows the I=0, $J^{PC}=1^{++}$ charmonium spectrum with hidden strange quarks. We identify 
the two low lying states represented by squares to be $\chi_{c1}(1P)$ and the level related to $X(3872)$. The 
remaining four levels are identified  with the $D_s\bar D_s^*$ and $J/\psi \phi$ scattering levels based
on overlap factors and behavior of the spectrum on omitting these operators. 
Thus we find no levels that could be related to the $Y(4140)$ or any other exotic structure below 4.2 GeV. 
Note that existence and the quantum numbers of most of the XYZ's are not yet settled from experiments. 
Therefore it is possible that the absence of additional levels in our studies is due to the fact that 
we explored the channel $J^P=1^{++}$ only.

\vspace{-0.2cm}
\section{Conclusions\label{sec:Conc}}
In this talk, we report the results from our lattice investigation of charmonium spectra with $J^{PC}=1^{++}$
and three different quark contents: $\bar cc \bar du$, $\bar cc(\bar uu+\bar dd)$ and $\bar cc \bar ss$, where 
the later two can mix with $\bar cc$. These calculations were performed on $N_f\!=\!2$ dynamical gauge 
configurations with $m_{\pi}\!\simeq \!266$ MeV. Using a large number of interpolators, including $\dadt{c}{q}{c}{q}$, 
$\dads{c}{q}{c}{q}$, $(\bar cq)_{1_c}(\bar qc)_{1_c}$, $(\bar cc)_{1_c}(\bar qq)_{1_c}$ and $(\bar cc)_{1_c}$,
we extract the spectra up to $4.2~$GeV. We identify and extract the lattice estimate for $\chi_{c1}$ and 
$X(3872)$, while all the remaining eigenstates are related to the expected two-meson scattering channels.
The $\bar cc$ Fock component in $X(3872)$ appears to be more important than the $\dad{c}{q}{c}{q}$, 
since we find a candidate for $X(3872)$ only if $\bar cc$ interpolating fields are used.
No additional levels were observed in the $I=1$ spectra with quark content $\bar cc \bar du$, which could have implied 
lattice candidate for charged or neutral $X(3872)$. Future simulations with broken isospin could be  crucial 
for this channel. We also do not find a candidate for $Y(4140)$ or any other exotic charmonium-like 
structure. Our search for the exotic states assumes an appearance of an additional energy eigenstate on the lattice, 
which is a typical manifestation for conventional hadrons.  Further analytic work is needed to establish whether this working 
assumption applies also for several coupled channels and all exotic structures of interest.

\acknowledgments\label{sec:Ackn}
We thank Anna Hasenfratz for providing the gauge configurations. We acknowledge the discussions 
with C. DeTar, L. Leskovec, D. Mohler, S. Ozaki and S. Sasaki. The calculations were performed on computing 
clusters at  the University of Graz (NAWI Graz), at the Vienna Scientific Cluster (VSC) and at Jozef Stefan 
Institute. This work is supported in part by the Austrian Science Fund FWF:I1313-N27 and by the Slovenian 
Research Agency ARRS project N1-0020. S.P. acknowledges support from U.S. Department of Energy contract 
DE-AC05-06OR23177, under which Jefferson Science Associates, LLC, manages and operates Jefferson Laboratory.

\end{document}